\documentclass{birkmult}

\usepackage{amsmath,amssymb,latexsym, amscd}
\usepackage{exscale, graphics, epsfig, epstopdf, esint, color}
\usepackage{eucal}

\usepackage[colorlinks]{hyperref}

\theoremstyle{definition}

\theoremstyle{remark}

\begin{document}
%
\firstpage{1}
\volume{XX}
\Copyrightyear{2018}
\DOI{XXX-XXXX}
\seriesextra{Just an add-on}
\seriesextraline{This is the Concrete Title of this Book\br H.E. R and S.T.C. W, Eds.}
%
%
%
%
\title[Stokes waves with vorticity]{Stokes waves in a constant vorticity flow}

\author[Dyachenko]{Sergey A. Dyachenko}
\email{sdyachen@math.uiuc.edu}
\author[Hur]{Vera Mikyoung Hur}
\email{verahur@math.uiuc.edu}
\address{%
Department of Mathematics, University of Illinois at Urbana-Champaign, Urbana, IL 61801 USA}
\thanks{This work was completed with the support of our \TeX-pert.}



\begin{abstract}
The Stokes wave problem in a constant vorticity flow is formulated via a conformal mapping as a modified Babenko equation. The associated linearized operator is self-adjoint, whereby efficiently solved by the Newton-conjugate gradient method. 
For strong positive vorticity, a fold develops in the wave speed versus amplitude plane, and a gap as the vorticity strength increases, bounded by two touching waves, whose profile contacts with itself, enclosing a bubble of air. More folds and gaps follow as the vorticity strength increases further. 
Touching waves at the beginnings of the lowest gaps tend to the limiting Crapper wave as the vorticity strength increases indefinitely, while a fluid disk in rigid body rotation at the ends of the gaps. Touching waves at the boundaries of higher gaps contain more fluid disks.
\end{abstract}

\maketitle


\section{Introduction}\label{sec:intro}

Stokes in his classical treatise~\cite{Stokes1847} (see also \cite{Stokes1880}) made formal but far-reaching considerations about periodic waves at the surface of an incompressible inviscid fluid in two dimensions, under the influence of gravity, which travel a long distance at a practically constant velocity without change of form. For instance, he observed that the crests become sharper and the troughs flatter as the amplitude increases, and that the `wave of greatest height' exhibits a $120^\circ$ corner at the crest. 
It would be impossible to give a complete account of Stokes waves here. We encourage the interested reader to some excellent surveys \cite{Toland1996, BT2003, Strauss2010}. 
We merely pause to remark that in an irrotational flow of infinite depth, notable recent advances were based on a formulation of the problem as a nonlinear pseudodifferential equation, involving the periodic Hilbert transform, originally due to Babenko~\cite{Babenko} (see also \cite{LH1978, Plotnikov1992, DKSZ1996}). For instance, \cite{BDT2002a, BDT2002b} (see also \cite{BT2003} and references therein) rigorously addressed the existence in-the-large, and \cite{DLK2016, Lushnikov2016, LDS2017} numerically approximated the wave of greatest height and revealed the structure of the complex singularities in great detail.  

The irrotational flow assumption is well justified in some circumstances. But rotational effects are significant in many others, for instance, for wind driven waves, waves in a shear flow, or waves near a ship or pier. 
Constant vorticity is of particular interest because it greatly simplifies the mathematics. 
Moreover, for short waves, compared with the characteristic lengthscale of vorticity, the vorticity at the fluid surface would be dominant. For long waves, compared with the fluid depth, the mean vorticity would be dominant (see the discussion in \cite{PTdS1988}). 

Simmen and Saffman~\cite{SS1985} and Teles da Silva and Peregrine~\cite{PTdS1988}, among others, employed a boundary integral method and numerically computed Stokes waves in a constant vorticity flow. Their results include overhanging profiles and interior stagnation points. To compare, a Stokes wave in an irrotational flow is necessarily the graph of a single valued function and each fluid particle must move at a velocity less than the wave speed. 

Recently, Constantin, Strauss and Varvaruca~\cite{CSV2016} used a conformal mapping, modified the Babenko equation and supplemented it with a scalar constraint, to permit constant vorticity and finite depth, and they rigorously established a global bifurcation result. The authors~\cite{DH1} rediscovered the modified Babenko equation and the scalar constraint, and numerically solved by means of the Newton-GMRES method (see also \cite{Choi2009, RMN2017}). More recently, the authors~\cite{DH2} eliminated the Bernoulli constant from the modified Babenko equation and, hence, the scalar constraint. The associated linearized operator is self-adjoint, whereby efficiently handled by means of the conjugate gradient method. Here we review the analytical formulation and numerical findings of \cite{DH1,DH2}.

For strong positive vorticity, the amplitude increases, decreases and increases during the continuation of the numerical solution. Namely, a {\em fold} develops in the wave speed versus amplitude plane, and it becomes larger as the vorticity strength increases. For nonpositive vorticity, on the other hand, the amplitude increases monotonically. For stronger positive vorticity, a {\em gap} develops in the wave speed versus amplitude plane, bounded by two {\em touching waves}, whose profile contacts with itself at the trough line, enclosing a bubble of air, and the gap becomes larger as the vorticity strength increases. By the way, the numerical method of \cite{SS1985, PTdS1988} and others diverges in a gap. More folds and gaps follow as the vorticity strength increases even further.
 
Moreover, touching waves at the beginnings of the lowest gaps tend to the {\em limiting Crapper wave} (see \cite{Crapper}) as the vorticity strength increases indefinitely --- a striking and surprising link between rotational and capillary effects --- while they tend to a {\em fluid disk in rigid body rotation} at the ends of the gaps. Touching waves at the beginnings of the second gaps tend to the circular vortex wave on top of the limiting Crapper wave in the infinite vorticity limit, and the circular vortex wave on top of itself at the ends of the gaps. Touching waves at the boundaries of higher gaps contain more circular vortices in like manner. 

\section{Formulation}\label{sec:formulation}

The water wave problem, in the simplest form, concerns the wave motion at the surface of an incompressible inviscid fluid in two dimensions, under the influence of gravity. Although an incompressible fluid may have variable density, we assume for simplicity that the density~$=1$. 
Suppose for definiteness that in Cartesian coordinates, the $x$ axis points in the direction of wave propagation and the $y$ axis vertically upward. Suppose that the fluid at time $t$ occupies a region in the $(x,y)$ plane, bounded above by a free surface $y=\eta(x,t)$ and below by the rigid bottom $y=-h$ for some constant~$h$, possibly infinite. Let 
\[
\varOmega(t)=\{(x,y)\in\mathbb{R}^2:-h<y<\eta(x,t)\}\quad\text{and}\quad\varGamma(t)=\{(x,\eta(x,t)):x\in\mathbb{R}\}.
\]
Let $\boldsymbol{u}=\boldsymbol{u}(x,y,t)$ denote the velocity of the fluid at the point $(x,y)$ and time $t$, and $P=P(x,y,t)$ the pressure. They satisfy the Euler equations for an incompressible fluid:
\begin{subequations}\label{E:ww0}
\begin{equation}\label{E:Euler}
\boldsymbol{u}_t+(\boldsymbol{u}\cdot\nabla)\boldsymbol{u}=-\nabla P+(0,-g)\quad\text{and}\quad
\nabla\cdot\boldsymbol{u}=0\quad\text{in $\varOmega(t)$},
\end{equation}
where $g$ is the constant due to gravitational acceleration. Let 
\[
\omega:=\nabla\times\boldsymbol{u}
\] 
denote constant vorticity. By the way, if the vorticity is constant throughout the fluid at the initial time then Kelvin's circulation theorem implies that it remains so at later times. We assume that there is no motion in the air and we neglect the effects of surface tension. The kinematic and dynamic conditions:
\begin{equation}\label{E:surface}
\eta_t+\boldsymbol{u}\cdot\nabla (\eta-y)=0\quad\text{and}\quad P=P_{atm}\quad\text{at $\varGamma(t)$}
\end{equation}
express that each fluid particle at the surface remains so at all times, and that the pressure there equals the constant atmospheric pressure $=P_{atm}$. In the finite depth, $h<\infty$, the kinematic condition states
\begin{equation}\label{E:bottom}
\boldsymbol{u}\cdot(0,-1)=0\quad\text{at $y=-h$}.
\end{equation}
\end{subequations}
We assume without loss of generality that the solutions of \eqref{E:ww0} are $2\pi$ periodic in the $x$ variable.

For any $h\in(0,\infty)$, $\omega\in \mathbb{R}$ and $c\in\mathbb{R}$, clearly,
\begin{equation}\label{def:shear}
\eta(x,t)=0,\quad\boldsymbol{u}(x,y,t)=(-\omega y-c,0)\quad\text{and}\quad P(x,y,t)=P_{atm}-gy
\end{equation}
solve \eqref{E:ww0}. We assume that some external effects such as wind produce such a constant vorticity flow and restrict the attention to waves propagating in \eqref{def:shear}. 

Let
\begin{equation}\label{def:Phi}
\boldsymbol{u}=(-\omega y-c,0)+\nabla\varPhi,
\end{equation}
whence $\Delta\varPhi=0$ in $\varOmega(t)$ by the latter equation of \eqref{E:Euler}. Naemly, $\varPhi$ is a velocity potential for the irrotational perturbation from \eqref{def:shear}. For nonconstant vorticity, $\varPhi$ is no longer viable to use. Let $\varPsi$ be a harmonic conjugate of $\varPhi$. Substituting \eqref{def:Phi} into the former equation of \eqref{E:Euler}, we make an explicit calculation to arrive at 
\begin{equation}\label{E:bernoulli}
\varPhi_t+\frac12(\varPhi_x^2+\varPhi_y^2)-(\omega y+c)\varPhi_x+\omega\varPsi+P-P_{atm}+gy=b(t)
\end{equation}
for some function $b(t)$. We substitute \eqref{def:Phi} into the other equations of \eqref{E:ww0}, likewise. The result becomes, by abuse of notation,
\begin{subequations}\label{E:ww}
\begin{align} 
&\Delta\varPhi=0 \quad &&\text{in $\varOmega(t)$}\\
&\eta_t+(\varPhi_x-\omega\eta-c)\eta_x=\varPhi_y &&\text{at $\varGamma(t)$}, \label{E:ww;K}\\
&\varPhi_t+\frac12|\nabla\varPhi|^2-(\omega\eta+c)\varPhi_x+\omega\varPsi+g\eta=0 &&\text{at $\varGamma(t)$},\label{E:ww;B}\\
&\varPhi_y=0 &&\text{at $y=-h$.}\label{E:ww;bottom}
\end{align}
By the way, since $\varPhi$ and $\varPsi$ are determined up to arbitrary functions of $t$, we may take without loss of generality that $b(t)=0$ at all times! 
In the infinite depth, $h=\infty$, we replace \eqref{E:ww;bottom} by
\begin{equation}\label{E:ww;infty}
\varPhi,\varPsi\to0\quad\text{as $y\to-\infty$}\quad\text{uniformly for $x\in\mathbb{R}$}.
\end{equation}
\end{subequations} 
See \cite{DH1,DH2}, for instance, for details.

\subsection{Reformulations in conformal coordinates}\label{sec:reformulation}

To proceed, we reformulate \eqref{E:ww} in conformal coordinates. Details may be found in \cite{DH1, DH2}. In what follows, we identify $\mathbb{R}^2$ with $\mathbb{C}$ whenever it is convenient to do so. 

Let
\begin{equation}\label{def:conformal}
z=z(w,t),\quad \text{where}\quad w=u+iv\quad\text{and}\quad z=x+iy,
\end{equation}
conformally map $\Sigma_{d}:=\{u+iv\in\mathbb{C}:-d<v<0\}$ of $2\pi$ period in the $u$ variable, to $\varOmega(t)$ of $2\pi$ period in the $x$ variable, for some $d$, possibly infinite. Let \eqref{def:conformal} extend to map $\{u+i0:u\in\mathbb{R}\}$ to $\varGamma(t)$, and $\{u-id:u\in\mathbb{R}\}$ to $\{x-ih:x\in\mathbb{R}\}$ if $d,h<\infty$, and $-i\infty$ to $-i\infty$ if $d,h=\infty$, where $d=\langle y\rangle+h$ (see \cite{DH1} for detail). Here and elsewhere,
\[
\langle f\rangle=\frac{1}{2\pi}\int_{-\pi}^{\pi} f(u)~du
\]
denotes the mean of a $2\pi$ periodic function $f$ over one period.

\subsubsection*{Periodic Hilbert transforms for a strip}

For $d$ in the range $(0,\infty)$, let 
\begin{align}
\mathcal{H}_de^{iku}=&-i\tanh(kd)e^{iku}&&\hspace*{-50pt}\text{for $k\in\mathbb{Z}$}\notag
\intertext{and}
\mathcal{T}_de^{iku}=&-i\coth(kd)e^{iku}&&\hspace*{-50pt}\text{for $k\neq0,\in\mathbb{Z}$}.\label{def:T}
\intertext{Let}
\mathcal{H}e^{iku}=&-i\,\text{sgn}(k)e^{iku}&&\hspace*{-50pt}\text{for $k\in\mathbb{Z}$}.\notag
\end{align}

When $d<\infty$, if $F$ is holomorphic in $\Sigma_d$ and $2\pi$ periodic in the $u$ variable and if $\text{Re}\,F(\cdot+i0)=f$ and $(\text{Re}\,F)_v(\cdot-id)=0$ then
\begin{equation}\label{E:1-iH}
F(\cdot+i0)=(1-i\mathcal{H}_d)f
\end{equation}
up to the addition by a purely imaginary constant. Namely, $1-i\mathcal{H}_d$ is the surface value of a periodic holomorphic function in a strip, the normal derivative of whose real part vanishes at the bottom. If $\text{Im}\,F(\cdot+i0)=f$ and $\text{Im}\,F(\cdot-id)=0$, and if $\langle f\rangle=0$, instead, then 
\begin{equation}\label{E:T+i}
F(\cdot+i0)=(\mathcal{T}_d+i)f
\end{equation}
up to the addition by a real constant. Namely, $\mathcal{T}_d+i$ is the surface value of a periodic holomorphic function in a strip, whose imaginary part is of mean zero at the surface and vanishes at the bottom. Moreover, when $d=\infty$, if $F$ is holomorphic in $\Sigma_\infty$ and $2\pi$ periodic in the $u$ variable and if $F$ vanishes sufficiently rapidly at $-i\infty$ then the real and imaginary parts of $F(\cdot+i0)$ are the periodic Hilbert transforms for each other (see \cite{Zygmund}, for instance).

\subsubsection*{Implicit form}

Note that $(x+iy)(u,t)$, $u\in\mathbb{R}$, makes a conformal parametrization of the fluid surface. In the finite depth, $d,h<\infty$, it follows from the Cauchy-Riemann equations and \eqref{def:T} that
\begin{equation}\label{E:Ty}
(x+iy)(u,t)=u+(\mathcal{T}_d+i)y(u,t).
\end{equation}
In the infinite depth, $d,h=\infty$, $\mathcal{H}$ replaces $\mathcal{T}_d$ (see \cite{DKSZ1996, DH2}, for instance). 

Moreover, let 
\[
(\phi+i\psi)(w,t)=(\varPhi+i\varPsi)(z(w,t),t)\quad\text{for $w\in\Sigma_d$}.
\] 
Namely, it is a conformal velocity potential for the irrotational perturbation from \eqref{def:shear}. In the finite or infinite depth, it follows from \eqref{E:1-iH} that
\begin{equation}\label{E:Hp}
(\phi+i\psi)(u,t)=(1-i\mathcal{H}_d)\phi(u,t)
\end{equation}
up to the addition by a purely imaginary constant.

In the finite depth, substituting \eqref{E:Ty} and \eqref{E:Hp} into \eqref{E:ww;K}-\eqref{E:ww;B}, we make an explicit calculation to arrive at
\begin{equation}\label{E:implicit}
\begin{aligned}
&(1+\mathcal{T}_dy_u)y_t-y_u\mathcal{T}_dy_t-\mathcal{H}_d\phi_u-(\omega y+c)y_u=0,\\
&((1+\mathcal{T}_dy_u)^2+y_u^2)(\phi_t+gy-\omega\mathcal{H}_d\phi)\\
&\quad-((1+\mathcal{T}_dy_u)\mathcal{T}_dy_t+y_uy_t)\phi_u
+(y_u\mathcal{T}_dy_t-(1+\mathcal{T}_dy_u)y_t)\mathcal{H}_d\phi_u \\
&\quad\quad+\frac12(\phi_u^2+(\mathcal{H}_d\phi_u)^2)
-(\omega y+c)((1+\mathcal{T}_dy_u)\phi_u-y_u\mathcal{H}_d\phi_u)=0.
\end{aligned}
\end{equation}
In the infinite depth, $\mathcal{H}$ replaces $\mathcal{H}_d$ and $\mathcal{T}_d$. See \cite{DH1}, for instance, for details. 

\subsubsection*{Explicit form}

In the finite depth, note that $z_t/z_u$ is holomorphic in $\Sigma_d$,
\[
\text{Im}\frac{z_t}{z_u}=\frac{\mathcal{H}_d\phi_u+(\omega y+c)y_u}{|z_u|^2}\quad\text{at $v=0$}
\]
by the former equation of \eqref{E:implicit}, and $\text{Im}(z_t/z_u)=0$ at $v=-d$ by \eqref{E:ww;bottom}. Note that $\langle \text{Im}(z_t/z_u)\rangle=0$ for any $v\in[-d,0]$ by the Cauchy-Riemann equations and \eqref{E:ww;bottom}. It then follows from \eqref{E:T+i} that 
\begin{equation}\label{E:zt/zu}
\frac{z_t}{z_u}=(\mathcal{T}_d+i)\Big(-\frac{(\mathcal{H}_d\phi+\frac12\omega y^2+cy)_u}{|z_u|^2}\Big)\quad\text{at $v=0$}.
\end{equation}
Moreover, note that $(\phi_u-i\mathcal{H}_d\phi_u)^2$ is the surface value of a holomorphic and $2\pi$ periodic function in $\Sigma_d$, the normal derivative of whose real part vanishes at the bottom. It then follows from \eqref{E:1-iH} and \eqref{def:T} that 
\begin{equation}\label{E:phi2}
\phi_u^2-(\mathcal{H}_d\phi_u)^2=-2\mathcal{T}_d(\phi_u\mathcal{H}_d\phi_u).
\end{equation}
We use \eqref{E:zt/zu} and \eqref{E:phi2}, and make a lengthy but explicit calculation to solve \eqref{E:implicit} as 
\begin{align}\label{E:explicit}
y_t=&(1+\mathcal{T}_dy_u+y_u\mathcal{T}_d)
\Big(\frac{\mathcal{H}_d\phi_u+(\omega y+c)y_u}{(1+\mathcal{T}_dy_u)^2+y_u^2}\Big), \notag \\
\phi_t=&-\phi_u\mathcal{T}_d
\Big(\frac{\mathcal{H}_d\phi_u+(\omega y+c)y_u}{(1+\mathcal{T}_dy_u)^2+y_u^2}\Big) \notag \\
&+\frac{1}{(1+\mathcal{T}_dy_u)^2+y_u^2}(\mathcal{T}_d(\phi_u\mathcal{H}_d\phi_u)
+(\omega y+c)(1+\mathcal{T}_dy_u)\phi_u)+\omega\mathcal{H}_d\phi-gy.
\end{align}
In the infinite depth, $\mathcal{H}$ replaces $\mathcal{H}_d$ and $\mathcal{T}_d$. See \cite{DH1}, for instance, for details.

\subsection{The Stokes wave problem in a constant vorticity flow}\label{sec:Stokes}

We turn the attention to the solutions of \eqref{E:explicit}, for which $y_t, \phi_t=0$. 

In the finite depth, substituting $y_t=0$ into the former equation of \eqref{E:explicit}, we arrive at
\begin{equation}\label{E:phi'}
\phi'=\mathcal{T}_d(\omega yy'+cy')\quad\text{at $v=0$}.
\end{equation}
Here and elsewhere, the prime denotes ordinary differentiation. 
Substituting $\phi_t=0$ into the latter equation of \eqref{E:explicit}, likewise, we use \eqref{E:phi'} and we make an explicit calculation to arrive at
\begin{equation}\label{E:Stokes}
(c+\omega y(1+\mathcal{T}_dy')-\omega\mathcal{T}_d(yy'))^2=(c^2-2gy)((1+\mathcal{T}_dy')^2+(y')^2).
\end{equation}
In the infinite depth, $\mathcal{H}$ replaces $\mathcal{T}_d$. 
If we were to take \eqref{E:bernoulli}, rather than \eqref{E:ww;B}, where $b=0$, then the result would become
\begin{equation}\label{E:DH1b}
(c+\omega y(1+\mathcal{T}_dy')-\omega\mathcal{T}_d(yy'))^2=(c^2+2b-2gy)((1+\mathcal{T}_dy')^2+(y')^2),
\end{equation}
and one must determine $b$ as part of the solution. See \cite{DH1}, for instance, for details.

\subsubsection*{The modified Babenko equation}

Unfortunately, \eqref{E:Stokes} or \eqref{E:DH1b} is not suitable for numerical solution, because one would have to work with rational functions of $y$. We reformulate \eqref{E:Stokes} as in a more convenient form. Details may be found in \cite{DH1, DH2}.

In the finite depth, we rearrange \eqref{E:Stokes} as
\begin{multline*}
(c-\omega\mathcal{T}_d(yy'))^2+2\omega y(c-\omega\mathcal{T}_d(yy'))(1+\mathcal{T}_dy')
-\omega^2y^2(y')^2 \\ =(c^2-2gy-\omega^2y^2)((1+\mathcal{T}_dy')^2+(y')^2).
\end{multline*}
Note that $(c-\omega(\mathcal{T}_d+i)(yy'))^2$ is the surface value of a holomorphic and $2\pi$ periodic function in $\Sigma_d$, whose imaginary part is of mean zero at the surface and vanishes at the bottom. Hence, so is
\begin{align*}
(&c^2-2gy-\omega^2y^2)((1+\mathcal{T}_dy')^2+(y')^2)
-2\omega y(c-\omega\mathcal{T}_d(yy'))(1+\mathcal{T}_dy'+iy') \\
&=((c^2-2gy-\omega^2y^2)(1+\mathcal{T}_dy'-iy')
-2\omega y(c-\omega\mathcal{T}_d(yy')))(1+\mathcal{T}_dy'+iy').
\end{align*}
Moreover, note that $1/(1+\mathcal{T}_dy'+iy')$ is the surface value of the holomorphic and $2\pi$ periodic function $=1/z_u$ in $\Sigma_d$, whose imaginary part is of mean zero at the surface and vanishes at the bottom. 
Hence, so is
\[
(c^2-2gy-\omega^2y^2)(1+\mathcal{T}_dy'-iy')-2\omega y(c-\omega\mathcal{T}_d(yy')).
\]
Therefore, it follows from \eqref{E:Ty} that
\[
(c^2-2gy-\omega^2y^2)(1+\mathcal{T}_dy')-2\omega y(c-\omega\mathcal{T}_d(yy'))
=-\mathcal{T}_d((c^2-2gy-\omega^2y^2)y')
\]
up to the addition by a real constant. Or, equivalently, 
\begin{multline}\label{E:babenko}
c^2\mathcal{T}_dy'-(g+c\omega)y-g(y\mathcal{T}_dy'+\mathcal{T}_d(yy'))\\
-\frac12\omega^2(y^2+\mathcal{T}_d(y^2y')+y^2\mathcal{T}_dy'-2y\mathcal{T}_d(yy'))=0
\end{multline}
and
\begin{equation}\label{E:mean}
g\langle y(1+\mathcal{T}_dy')\rangle+c\omega\langle y\rangle+\frac12\omega^2\langle y^2\rangle=0.
\end{equation}
Indeed, $\langle \mathcal{T}_df'\rangle=0$ for any function $f$ by \eqref{def:T} and 
\[
\langle y^2\mathcal{T}_dy'\rangle =\frac{1}{2\pi}\int^{\pi}_{-\pi} y^2\mathcal{T}_dy'~du
=-\frac{1}{2\pi}\int^{\pi}_{-\pi} y\mathcal{T}_d(y^2)'~du=-\langle 2y\mathcal{T}_d(yy')\rangle.
\]
In the infinite depth, $\mathcal{H}$ replaces $\mathcal{T}_d$. Conversely, a solution of \eqref{E:babenko}-\eqref{E:mean} gives rise to a traveling wave of \eqref{E:ww} and, hence, \eqref{E:ww0}, provided that 
\begin{subequations}\label{C:limiting}
\begin{gather}
u\mapsto (u+\mathcal{T}_dy(u), y(u)), u\in\mathbb{R}, \text{is injective} \label{C:touching}
\intertext{and}
((1+\mathcal{T}_dy_u)^2+y_u^2)(u)\neq0\quad\text{for any $u\in\mathbb{R}$}. \label{C:extreme}
\end{gather} 
\end{subequations}
See \cite{DH1,DH2}, for instance, for details.
The Stokes wave problem in a constant vorticity flow is to find $\omega\in\mathbb{R}$, $d\in(0,\infty]$, $c\in\mathbb{R}$ an a $2\pi$ periodic function $y$, satisfying \eqref{C:limiting}, which together solve \eqref{E:babenko}-\eqref{E:mean}. 
In what follows, we assume that $y$ is even (see \cite{Hur2007}, for instance, for arbitrary vorticity).

In an irrotational flow of infinite depth, $\omega=0$ and $d=\infty$, \eqref{E:babenko} and \eqref{E:mean} simplify to
\begin{equation}\label{E:babenko0}
c^2\mathcal{H}y'-gy-g(y\mathcal{H}y'+\mathcal{H}(yy'))=0
\end{equation}
and $\langle y(1+\mathcal{T}_dy')\rangle=0$. Longuet-Higgins~\cite{LH1978} discovered a set of identities among the Fourier coefficients of a Stokes wave, which Babenko~\cite{Babenko} rediscovered in the form of \eqref{E:babenko0} and, independently, \cite{Plotnikov1992, DKSZ1996} among others. One may regard \eqref{E:babenko}-\eqref{E:mean} as the modified Babenko equation, permitting constant vorticity and finite depth. 

If we were to take \eqref{E:bernoulli}, rather than \eqref{E:ww;B}, where $b=0$, then \eqref{E:babenko} would become
\begin{equation}\label{E:DH1y}
\begin{aligned}
(c^2+2b)\mathcal{T}_dy'-&(g+c\omega)y-g(y\mathcal{T}_dy'+\mathcal{T}_d(yy'))\\
-&\frac12\omega^2(y^2+y^2\mathcal{T}_dy'+\mathcal{T}_d(y^2y')-2y\mathcal{T}_d(yy'))=0,
\end{aligned}
\end{equation}
which is supplemented with  
\begin{equation}\label{E:DH1-mean}
\langle(c+\omega y(1+\mathcal{T}_dy')-\omega\mathcal{T}_d(yy'))^2\rangle
=\langle (c^2+2b-2gy)((1+\mathcal{T}_dy')^2+(y')^2)\rangle.
\end{equation}
This is what \cite{CSV2016, DH1} derived. 

\section{Numerical method}\label{sec:numerical}

We write \eqref{E:babenko} in the operator form as $\mathcal{G}(y; c, \omega, d)=0$ and solve it iteratively using the Newton method. Let $y^{(n+1)}=y^{(n)}+\delta y^{(n)}$, $n=0,1,2,\dots$, where $y^{(0)}$ is an initial guess, to be supplied (see \cite{DH1, DH2}, for instance), and $\delta y^{(n)}$ solves 
\begin{equation}\label{E:dF}
\delta \mathcal{G}(y^{(n)};c,\omega,d)\delta y^{(n)}=-\mathcal{G}(y^{(n)};c,\omega,d),
\end{equation}
$\delta \mathcal{G}(y^{(n)};c,\omega,d)$ is the linearization of $\mathcal{G}(y;c,\omega,d)$ with respect to $y$ and evaluated at $y=y^{(n)}$. 

We exploit an auxiliary conformal mapping, involving Jacobi elliptic functions (see \cite{DH2} and references therein), and take efficient, albeit highly nonuniform, grid points in $u\in[-\pi,\pi]$. We approximate $y^{(n)}$ by a discrete Fourier transform and numerically evaluate $y^{(n)}$, $\mathcal{T}_dy^{(n)}$ and others using a fast Fourier transform. Since
\begin{align*}
\delta\mathcal{G}(y;c,\omega,d)\delta y=&c^2\mathcal{T}_d(\delta y)'-(g+c\omega)\delta y
-g(\delta y\mathcal{T}_dy'+y\mathcal{T}_d(\delta y)'+\mathcal{T}_d(y\delta y)') \\
&-\frac12\omega^2(2y\delta y+\mathcal{T}_d(y^2\delta y)'-[2y\delta y,y]+[y^2,\delta y]),
\end{align*}
where $[f_1,f_2]=f_1\mathcal{T}_df_2-f_2\mathcal{T}_df_1$, is self-adjoint, we solve \eqref{E:dF} using the conjugate gradient (CG) method. 
We employ \eqref{E:mean} to determine the zeroth Fourier coefficient.
Once we arrive at a convergent solution, we continue it along in the parameters. See \cite{DH2}, for instance, for details. 

If we were to take \eqref{E:DH1y}-\eqref{E:DH1-mean}, rather than \eqref{E:babenko}, where $b=0$, then the associated linearized operator includes
\begin{align*}
(\delta y,\delta b)\mapsto &(c^2+2b)\mathcal{T}_d(\delta y)'\hspace*{-2pt}+\hspace*{-2pt}2\delta b\mathcal{T}_dy'
\hspace*{-2pt}-\hspace*{-2pt}(g+c\omega)\delta y 
\hspace*{-2pt}-\hspace*{-2pt}g(\delta y\mathcal{T}_dy'+y\mathcal{T}_d(\delta y)'+\mathcal{T}_d(y\delta y)')\\
&-\frac12\omega^2(2y\delta y+\mathcal{T}_d(y^2\delta y)'-[2y\delta y,y]+[y^2,\delta y]),
\end{align*}
which is {\em not} self-adjoint, whence the CG or conjugate residual method may not apply. The authors~\cite{DH1} used the generalized minimal residual (\mbox{GMRES}) method and achieved some success. But it would take too much time to accurately resolve a numerical solution when it requires excessively many grid points. The CG method is more powerful than the GMRES for self-adjoint equations, and it leads to new findings, which we discuss promptly.

\section{Results}\label{sec:results}

Summarized below are the key findings of \cite{DH1,DH2}. 

We take without loss of generality that $c$ is positive, and allow $\omega$ positive or negative, representing waves propagating upstream or downstream, respectively (see the discussion in \cite{PTdS1988}). 

We take for simplicity that $g=1$ and $d=\infty$. By the way, the effects of finite depth change the amplitude of a Stokes waves and others, but they are insignificant otherwise (see \cite{DH1}, for instance). 

In what follows, the steepness $s$ measures the crest-to-trough wave height divided by the period~$=~2\pi$. 

\subsection{Folds and gaps}\label{sec:fold/gap}

\begin{figure}[h]
\includegraphics[scale=1.0]{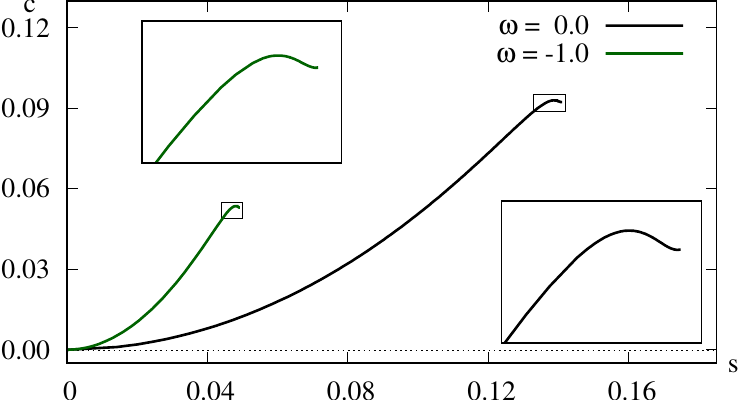}
\includegraphics[scale=1.0]{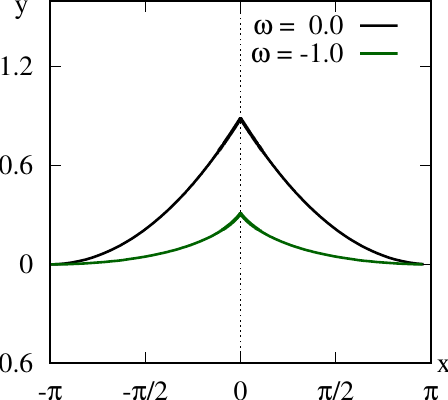}
\caption{On the left, wave speed vs. steepness for $\omega=0$ and~$-1$. Insets are closeups near the endpoints of the continuation of the numerical solution. On the right, the profiles of almost extreme waves.}
\label{fig:w=0,-1}
\end{figure}

For zero and negative constant vorticity, for instance, for $\omega=0$ and $-1$, the left panel of Figure~\ref{fig:w=0,-1} collects the wave speed versus steepness from the continuation of the numerical solution. For $\omega=0$, Longuet-Higgins and Fox~\cite{LHFox1978}, among others, predicted that $c$ oscillates infinitely many times whereas $s$ increases monotonically toward the wave of greatest height or the extreme wave, whose profile exhibits a $120^\circ$ corner at the crest. Numerical computations (see \cite{DLK2016, LDS2017}, for instance, and references therein) bear it out. The insets reproduce the well-known result and suggest likewise when $\omega=-1$. 

The right panel displays the profiles of almost extreme waves, in the $(x,y)$ plane in the range $x\in[-\pi,\pi]$. Troughs are at $y=0$. Note that the steepness when $\omega=-1$ is noticeably less than $\omega=0$. 

\begin{figure}[h]
\includegraphics[scale=1.0]{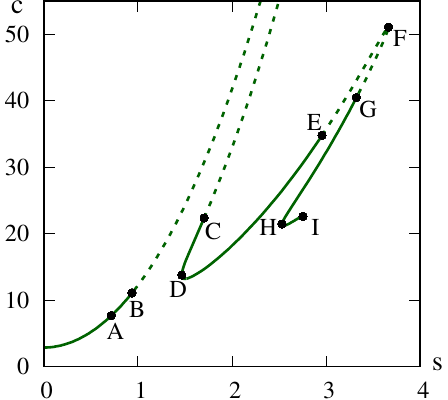}
\includegraphics[scale=1.0]{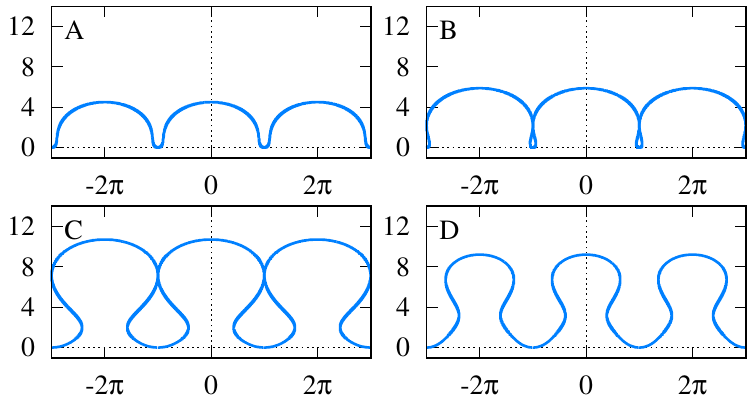}
\includegraphics[scale=1.0]{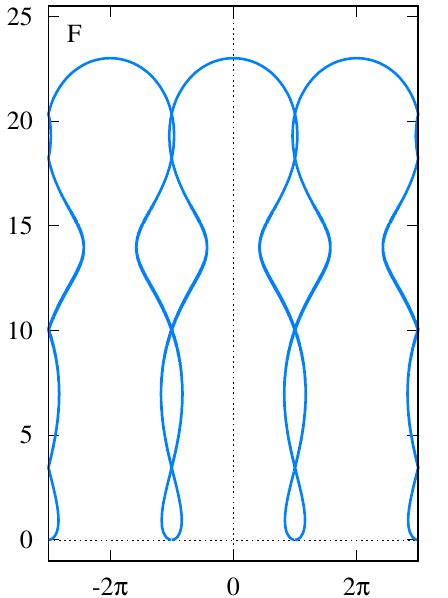}
\includegraphics[scale=1.0]{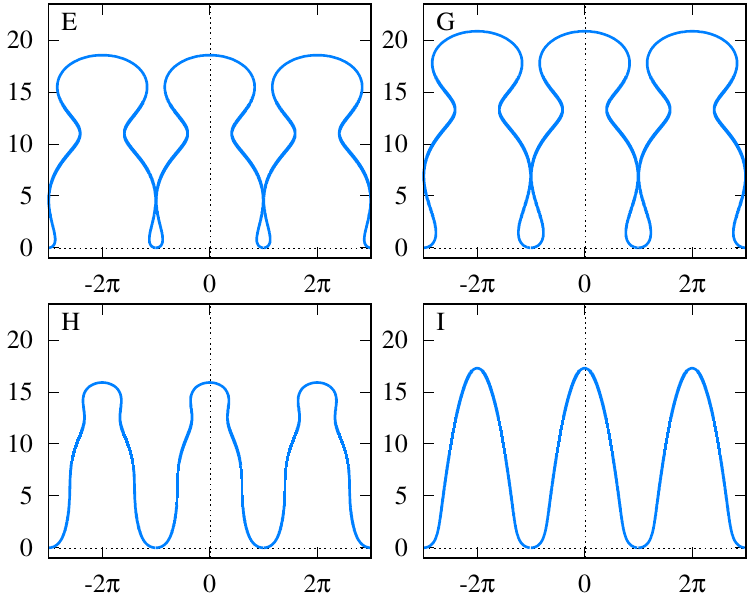}
\caption{For $\omega=2.5$. Clockwise from upper left: wave speed vs. steepness; the profiles of eight solutions, labelled by $A$ to $E$, and $G$ to $I$; the profile of an unphysical solution labelled by~$F$.}
\label{fig:w=2.5}
\end{figure}

For a large value of positive constant vorticity, for instance, for $\omega=2.5$, Figure~\ref{fig:w=2.5} includes the wave speed versus steepness and the profiles at the indicated points along the $c=c(s)$ curve, in the $(x,y)$ plane, where $x\in[-3\pi,3\pi]$. Troughs are at $y=0$. The upper left panel reveals that $s$ increases and decreases from $s=0$ to wave $D$. Namely, a {\em fold} develops in the $c=c(s)$ curve. For $s$ small, for instance, for wave~$A$, the profile is single valued. But we observe that the profile becomes more rounded as $s$ increases along the fold, so that overhanging waves appear, whose profile is no longer single valued. Moreover, we arrive at a {\em touching wave}, whose profile becomes vertical and contacts with itself somewhere the trough line, whereby enclosing a bubble of air. Wave $B$ is an almost touching wave. 

Past the touching wave, a numerical solution is unphysical because \eqref{C:touching} no longer holds true (see \cite{DH1, DH2} for examples). Moreover, we observe that the profile becomes less rounded as $s$ decreases along the fold, so that we arrive at another touching wave; past the touching wave, a numerical solution is physical. Wave~$C$ is an almost touching wave and wave $D$ is physical. Together, a {\em gap} develops in the $c=c(s)$ curve, consisting of unphysical numerical solutions and bounded by two touching waves. We remark that wave $C$ encloses a larger bubble of air than wave $B$. 

Past the end of the fold, interestingly, the upper left panel reveals another fold and another gap. The steepness increases from waves $D$ to $F$, and decreases from waves $F$ to $H$. Waves $E$ and $G$ are almost touching waves and numerical solutions between are unphysical. For instance, for wave $F$, the profile intersects itself and the fluid region overlaps itself. 

Past the end of the second fold, we observe that $s$ increases monotonically, although $c$ oscillates (see \cite{DH2}, for instance, for details), like when $\omega=0$; moreover, overhanging profiles disappear as $s$ increases and the crests become sharper, like when $\omega=0$. Therefore, we may claim that an {\em extreme wave} ultimately appears, whose profile exhibits a sharp corner at the crest. Wave $I$ is an almost extreme wave. One may not continue the numerical solution past the extreme wave because \eqref{C:extreme} would no longer hold true. 

\begin{figure}[h]
\includegraphics[scale=1.0]{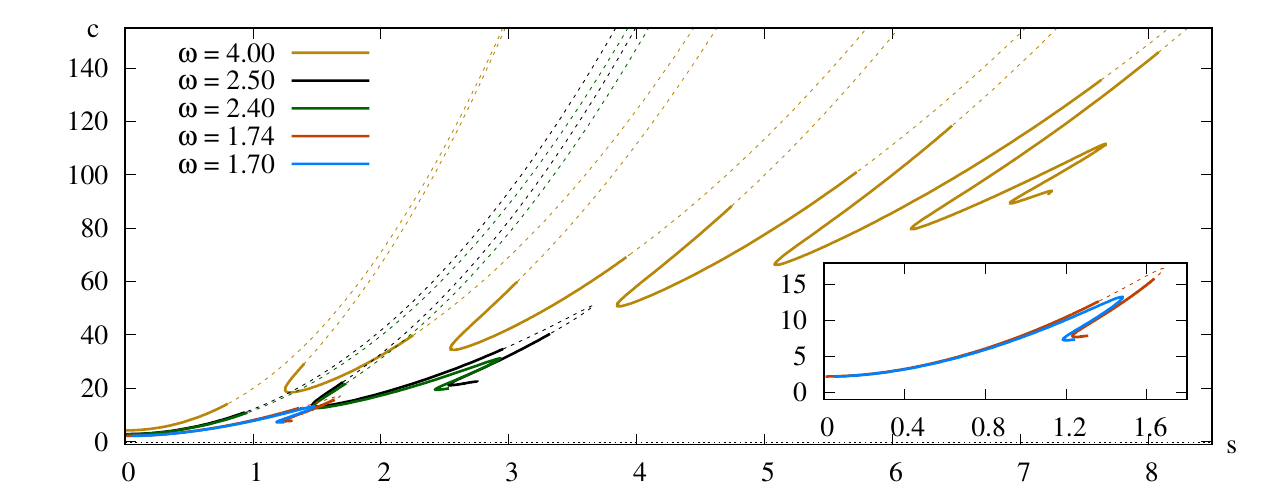}
\caption{Wave speed vs. steepness for five values of positive constant vorticity. Solid curves for physical solutions and dashed curves unphysical. The inset distinguishes the lowest fold and gap.}
\label{fig:c=c(s)}
\end{figure}

Figure~\ref{fig:c=c(s)} includes the wave speed versus steepness for several values of positive constant vorticity. For zero vorticity, one predicts that $c$ experiences infinitely many oscillations whereas $s$ increases monotonically (see \cite{LHFox1978}, for instance). For negative constant vorticity, numerical computations (see \cite{SS1985, PTdS1988, DH1}, among others) suggest that the crests become sharper and lower. Figure~\ref{fig:w=0,-1} bears it out.

For positive constant vorticity, for instance, for $\omega=1.7$, on the other hand, Figure~\ref{fig:c=c(s)} reveals that the lowest oscillation of $c$ deforms into a fold. Consequently, there correspond two or three solutions for some values of $s$. Moreover, the extreme wave seems not the wave of greatest height. We observe that the fold becomes larger in size as $\omega$ increases. For a larger value of the vorticity, for instance, for $\omega=1.74$, the figure reveals that part of the fold transforms into a gap. 
We observe that the gap becomes larger in size as $\omega$ increases. See \cite{DH1}, for instance, for details.

Moreover, for $\omega=2.4$, Figure~\ref{fig:c=c(s)} reveals that the second oscillation of $c$ deforms into another fold, and we observe that the second fold becomes larger in size as $\omega$ increases. For $\omega=2.5$, part of the second fold transforms into another gap, and we observe that the second gap becomes larger in size as $\omega$ increases. We merely pause to remark that the numerical method of \cite{SS1985, PTdS1988} and others diverges in a gap and is incapable of locating a second gap. The numerical method of \cite{DH1} converges in a gap, but it would take too much time to accurately resolve a numerical solution along a second fold.

We take matters further and claim that higher folds and higher gaps develop in like manner as $\omega>0$ increases. For instance, for $\omega=4$, Figure~\ref{fig:c=c(s)} reveals five folds and five gaps! Moreover, we claim that past all the folds, the steepness increases monotonically toward an extreme wave. Numerical computations (see \cite{DH2}, for instance) suggest that the extreme profile is single valued and exhibits a $120^\circ$ corner at the crest, regardless of the value of the vorticity. 

\subsection{Touching waves in the infinite vorticity limit}\label{sec:touching}

\begin{figure}[h]
\centerline{
\includegraphics[scale=1]{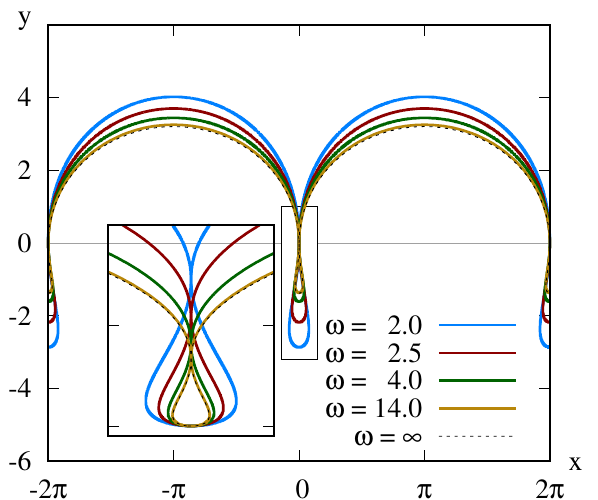}
\includegraphics[scale=1]{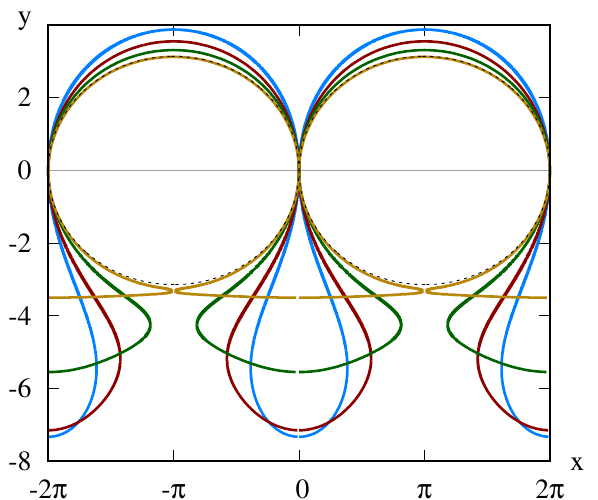}}
\caption{On the left, touching waves at the beginnings of the lowest gaps for four values of vorticity. The dashed curved line is the limiting Crapper wave. On the right, touching waves at the ends of the gaps. The dashed curved line is a circle.}
\label{fig:touching1}
\end{figure}

The left panel of Figure~\ref{fig:touching1} displays the profiles of almost touching waves near the beginnings of the lowest gaps, and the right panel near the ends of the gaps, for four values of positive constant vorticity, in the $(x,y)$ plane in the range $x\in[-2\pi,2\pi]$. Touching is at $y=0$. The profiles on the left resemble that in \cite[Figure~4(b)]{VB1996}. 

At the beginnings of the gaps, we observe that $s$ decreases monotonically toward $\approx0.73$ as $\omega\to\infty$ (see \cite{DH1}, for instance). Crapper~\cite{Crapper} derived a remarkable formula of periodic capillary waves (in the absence of gravitational effects) in an irrotational flow of infinite depth, and calculated that $s\approx0.73$ for the wave of greatest height. Moreover, the left panel reveals that, for instance, for $\omega=14$, the profile of an almost touching wave is in excellent agreement with the limiting Crapper wave. Therefore, we may claim that touching waves at the beginnings of the lowest gaps tend to the {\em limiting Crapper wave} as the value of positive constant vorticity increases indefinitely. It reveals a striking and surprising link between positive constant vorticity and capillarity!

At the ends of the gaps, on the other hand, we observe that $s\to1$ as $\omega\to\infty$ (see \cite{DH1}, for instance). Teles da Silva and Peregrine~\cite{PTdS1988}, among others, numerically computed periodic waves in a constant vorticity flow in the absence of gravitational effects, and argued that a limiting wave has a circular shape made up of fluid in rigid body rotation (see also \cite{VB1996}). Moreover, the right panel reveals that, for instance, for $\omega=14$, the profile of an almost touching wave is nearly circular. Therefore, we may claim that touching waves at the ends of the lowest gaps tend to a {\em fluid disk in rigid body rotation} in the infinite vorticity limit. 
s
It is interesting to analytically explain the limiting Crapper wave and the circular vortex wave in the infinite vorticity limit.

\begin{figure}[h]
\centerline{
\includegraphics[scale=1]{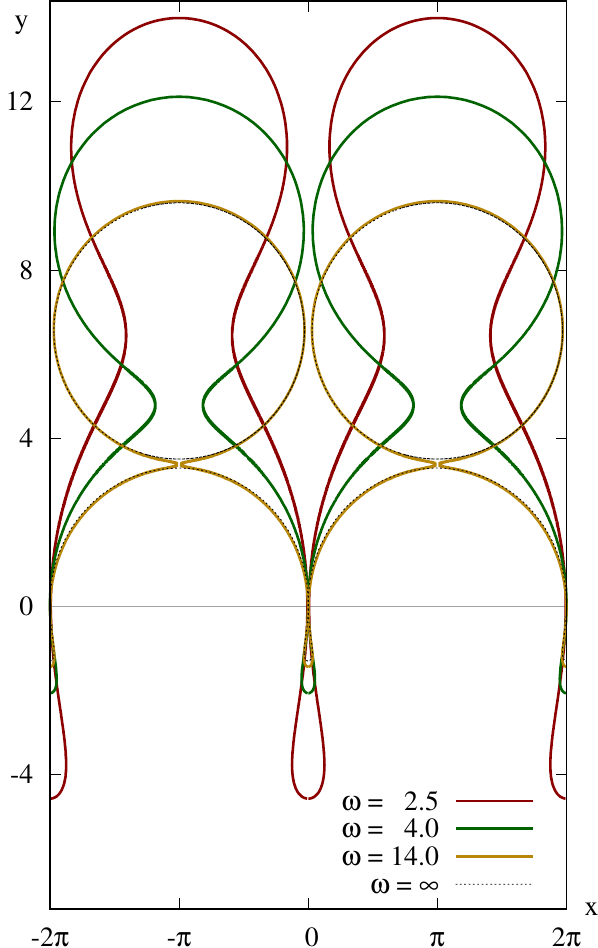}
\includegraphics[scale=1]{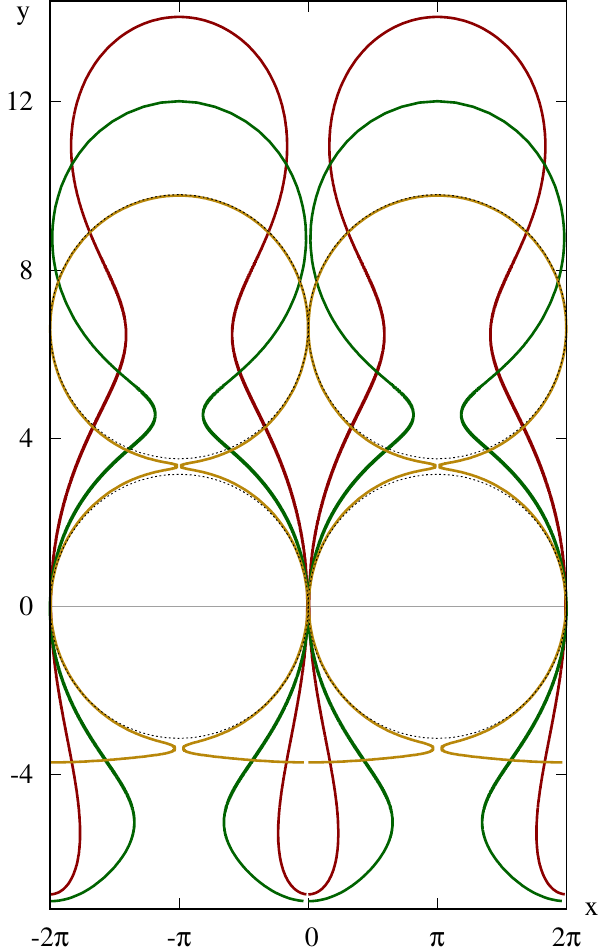}}
\caption{On the left, touching waves at the beginnings of the second gaps for three values of vorticity (solid) and the circular vortex wave on top of the limiting Crapper wave (dashed). On the right, touching waves at the ends of the gaps (solid) and the circular vortex wave on top of itself (dashed).}
\label{fig:touching2}
\end{figure}

Moreover in the left panel of Figure~\ref{fig:touching2} are the profiles of almost touching waves near the beginnings of the second gaps, and the right panel near the ends of the gaps, for three values of positive constant vorticity, in the $(x,y)$ plane, where $x\in[-2\pi,2\pi]$. The profile on the left for $\omega=14$ resembles that in \cite[Figure~5(c)]{VB1996}, and the profile on the right resembles \cite[Figure~6]{VB1996}. We may claim that touching waves at the beginnings of the second gaps tend to the circular vortex on top of the limiting Crapper wave as the value of positive constant vorticity increases indefinitely, whereas the circular vortex wave on top of itself at the ends of the gaps. 

We take matters further and claim that touching waves at the boundaries of higher gaps accommodate more circular vortices in like manner. See \cite{DH2}, for instance, for a profile nearly enclosing five circular vortices!



\subsection*{Acknowledgment}
VMH is supported by the US National Science Foundation under the Faculty Early Career Development (CAREER) Award DMS-1352597, and SD is supported by the National Science Foundation under DMS-1716822. VMH is grateful to the Erwin Schr\"odinger International Institute for Mathematics and Physics for its hospitality during the workshop Nonlinear Water Waves. 


\end{document}